\documentstyle[preprint,prb,aps,eqsecnum,psfig,amssymb]{revtex}
\tighten

\begin{document}
\draft

\title{Energy spectra, wavefunctions and quantum diffusion for
       quasiperiodic systems}
\author{H.~Q.~Yuan,$^{1,2,3}$ U.~Grimm,$^{1}$ 
        P.~Repetowicz,$^{1}$ and M.~Schreiber$^{1}$}
\address{$^{1}$Institut f\"ur Physik, 
         Technische Universit\"at Chemnitz, 
         09107 Chemnitz, Germany\\
         $^{2}$Max-Planck-Institut f\"{u}r Chemische Physik fester Stoffe,
         N\"{o}thnitzer Str.\ 40, 01187  Dresden,
         Germany\\
         $^{3}$Department of Physics, Xiangtan University,
         Xiangtan 411105, P.~R.~China}
\date{\today}
\maketitle

\begin{abstract}%
We study energy spectra, eigenstates and quantum diffusion for one-
and two-dimensional quasiperiodic tight-binding models. As our
one-dimensional model system we choose the silver mean or
``octonacci'' chain. The two-dimensional labyrinth tiling, which is
related to the octagonal tiling, is derived from a product of two
octonacci chains. This makes it possible to treat rather large systems
numerically. For the octonacci chain, one finds singular continuous
energy spectra and critical eigenstates which is the typical behaviour
for one-dimensional Schr\"{o}dinger operators based on substitution
sequences. The energy spectra for the labyrinth tiling can, depending
on the strength of the quasiperiodic modulation, be either band-like
or fractal-like. However, the eigenstates are multifractal. The
temporal spreading of a wavepacket is described in terms of the
autocorrelation function $C(t)$ and the mean square displacement
$d(t)$. In all cases, we observe power laws $C(t)\sim t^{-\delta}$ and
$d(t)\sim t^{\beta}$. For the octonacci chain, $0<\delta<1$, whereas
for the labyrinth tiling a crossover is observed from $\delta=1$ to
$0<\delta<1$ with increasing modulation strength. Corresponding to the
multifractal eigenstates, we obtain anomalous diffusion with
$0<\beta<1$ for both systems.  Moreover, we find that the behaviour of
$C(t)$ and $d(t)$ is independent of the shape and the location of the
initial wavepacket. We use our results to check several relations between 
the diffusion exponent $\beta$ and the fractal dimensions of energy 
spectra and eigenstates that were proposed in the literature. 
\end{abstract}


\narrowtext

\section{Introduction}
\label{sec1}

Since the pioneering discoveries of quasicrystals with
icosahedral,\cite{SBGC} dodecagonal,\cite{INF} decagonal,\cite{Ben}
and octagonal\cite{WCK} symmetry, electronic transport phenomena
arguably belong to the most celebrated and intriguing physical
properties of these intermetallic alloys.\cite{Poo,Ber,CR} For instance,
the electric conductivity of icosahedral quasicrystals decreases
strongly with decreasing temperature and with improving structural
quality of the sample, and anomalous transport behaviour is also
observed in other quantities such as thermopower or
magnetoconductance. Stimulated by the experimental results, a lot of
effort has been spent towards a better theoretical understanding of
the transport phenomena in quasicrystalline
materials.\cite{CR,SGD,1D,BYP,TM,CC,2D,RGT,JR,HJ,CG,SD,HHTM,BCV,IG,FP,RKST}
This is also of interest from the theoretical or mathematical point of
view, because quasicrystals as ordered aperiodic structures are
intermediate between periodically ordered crystals and short-range
ordered amorphous solids. In particular, the anomalous diffusion of
wave packets in quasiperiodic systems has attracted wide
interest.\cite{RGT,JR,HJ,CG,SD,HHTM,BCV,IG,FP,RKST}

Multifractal eigenstates --- neither extended over the system, nor
exponentially localized --- exist at the metal-insulator transition of
the Anderson model of localization.\cite{MHT,ARM} In tight-binding
models of quasicrystals, this kind of eigenstates has also been
revealed.\cite{1D,BYP,TM,CC,RGS} Generally, the energy spectra of
one-dimensional (1D) quasicrystals are singular continuous.\cite{1D}
However, in higher-dimensional cases, the energy spectra can be
band-like with finite measure, fractal-like with zero band width or a
mixture of partly band-like and partly fractal-like
character.\cite{CC,2D}

The diffusion properties of quasicrystals are associated with the complex
eigenstates and energy spectra stated above.\cite{IG,FP,RKST} To describe the
diffusion of a wave packet initially localized at some site $n_{0}$, one
usually discusses the temporal autocorrelation
function\cite{RGT,JR,HJ,CG,SD,TT,BL}
\begin{equation}
C(t) = \frac{1}{t}\int\limits_{0}^{t}|\Psi_{n_{0}}(t^{\prime})|^{2}\,
dt^{\prime}
\end{equation}
or the mean square displacement\cite{SD,HHTM,BCV,TT}    
\begin{equation}
d(t) = \left(\sum\limits_{n}|{\bf r}_{n}- {\bf r}_{n_{0}}|^{2}\,
             |\Psi_{n}(t)|^{2}\right)^{1/2}
\end{equation}
where $\Psi_{n}(t)$ is the amplitude of the wavefunction at time $t$
at the $n$th site which is located at the position ${\bf r}_{n}$ in
space.  Apparently, $C(t)$ is the time-averaged probability of a wave
packet staying at the initial site at time $t$, and $d(t)$ determines
the spreading of the width of a wave packet.

Generally, one finds $ C(t)\sim t^{-\delta}, d(t)\sim t^{\beta}$ with
$0<\delta <1$ and $ 0<\beta<1 $ for 1D quasiperiodic
systems.\cite{RGT,JR,CG,SD,HHTM} For higher-dimensional cases, no
general results are available. Zhong {\it et al.}\/ \cite{JR} observed
a transition of $C(t)$ with the increase of the quasiperiodic
modulation strength in simple higher-dimensional Fibonacci
lattices. For small patches of the octagonal tiling, Passaro {\it et
al.}\/ \cite{BCV} found $d(t)\sim t^\beta$ with $0<\beta<1$ even for
the case of a band-like spectrum. However, one of us obtained
$C(t)\sim t^{-1}$ for this case after analyzing the long-time
behaviour.\cite{HJ} In fact, it is quite difficult to derive the exact
long-time behaviour of $C(t)$ and $d(t)$ from the investigation of
rather small systems. Therefore, a study of a large higher-dimensional
quasiperiodic system will be significant. In this paper, we will
mainly discuss the diffusion properties on a 2D quasiperiodic tiling
related to the octagonal quasicrystals. The tiling is based on the
octonacci chain and thus permits us to study large systems.

Recent investigations show that the diffusion properties are connected
with the multifractality of eigenstates and energy
spectra.\cite{RGT,JR,HJ,CG,SD,HHTM,BCV,IG,FP,RKST,TT,BL,zhong} It can be
rigorously proven that the exponent $\delta$ ruling the decay of the
autocorrelation function $C(t)$ equals the correlation dimension
$D_{2}$ of the local spectral measure associated with the initial
site.\cite{RGT,JR} In 1D quasiperiodic systems, Guarneri\cite{IG}
analytically deduced that $\beta \geq D_{1}$, where $D_{1}$ is the
information dimension of the spectral measure. More recently,
Ketzmerick {\it et al.}\/ \cite{RKST} argued that $\beta$ is also
related to the multifractal properties of the eigenstates. We shall
address the question whether these relations exist in different
quasiperiodic systems, especially in higher-dimensional cases.

This paper is organized as follows. In the next section, we describe
the construction of the labyrinth tiling and its properties that are
relevant for our analysis.  Afterwards, in Sec.~\ref{sec2b}, we
consider a tight-binding model on the labyrinth tiling and express the
eigenstates and eigenvalues in terms of eigenstates and eigenvalues of
a tight-binding Hamiltonian on the octonacci chain. In Sec.~\ref{sec3}
we show the energy spectra and multifractal eigenstates for both these
systems.  Sec.~\ref{sec4} describes the diffusion properties of the
octonacci chain. The diffusion properties of the labyrinth tiling will
be emphasized in Sec.~\ref{sec5}. In Sec.~\ref{sec6} we discuss the
fractal dimensions of eigenstates and eigenspectra and their relation
to the diffusion properties. Finally, we conclude in Sec.~\ref{sec7}.

\section{The labyrinth tiling}
\label{sec2a}

The labyrinth tiling\cite{SMS,CC} can be considered as a subset of the
octagonal quasiperiodic tiling,\cite{octagonal} and vice
versa.\cite{CC} One can build it directly from the octonacci chain. In
order to construct the labyrinth tiling, we introduce the octonacci
sequence which can be produced by iterating the inflation rule
\begin{equation}
\varrho:\;\begin{array}{lcl} S & \rightarrow & L \\
                             L & \rightarrow & LSL \end{array}
\label{eq:subst}
\end{equation}
on the initial word $w_{0}=S$. The number of letters $g_{m}$
in the $m$th iterate $w_{m}=\varrho^{m}(w_{0})$ satisfies the recursion
\begin{equation} 
g_{m}=2g_{m-1}+g_{m-2},\qquad g_{0}=g_{1}=1.
\label{eq:gm}
\end{equation}
The numbers of letters $L$ and $S$ in $w_{m}$ are given by 
$f_{m}$ and $f_{m-1}$, respectively, which fulfill the
same recursion relation with a different initial condition
\begin{equation} 
f_{m}=2f_{m-1}+f_{m-2},\qquad f_{0}=0,\quad f_{1}=1,
\label{eq:fm}
\end{equation}
such that $g_{m}=f_{m}+f_{m-1}$. Their ratio in the limit sequence 
$w_{\infty}$
\begin{equation}
\lim_{m\rightarrow \infty} \frac{f_{m}}{f_{m-1}}=
\lim_{m\rightarrow \infty} \frac{g_{m}}{g_{m-1}}=\lambda
\end{equation}
is given by the silver mean $\lambda=1+\sqrt{2}$ which is a root of
the quadratic equation $\lambda^{2}=2\lambda+1$. As can been seen from
Eq.~(\ref{eq:gm}), $g_m$ is odd for all $m$.

Associating with the letters $S$ and $L$ intervals of length $1$ and
$\lambda$, respectively, one obtains a linear chain ${\cal C}_{m}$ of
$N_{m}=g_{m}+1$ sites, which is known as octonacci or silver mean
chain.\cite{CC} We note that all words $w_{m}$ obtained from the
substitution rule (\ref{eq:subst}) are palindromes, thus the resulting
chains are symmetric under reflection.  

\begin{figure} 
\centerline{\psfig{figure=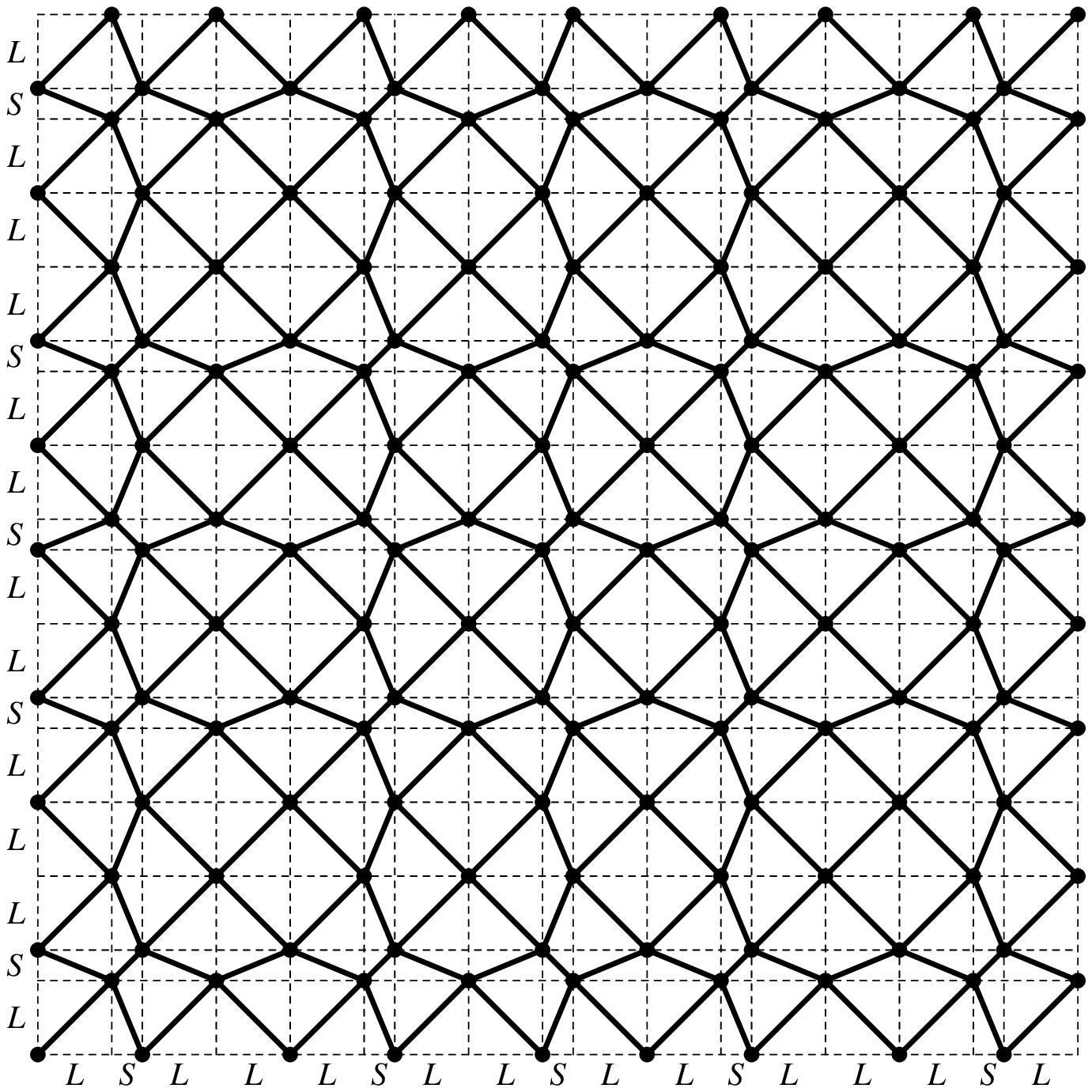,width=\columnwidth}}
 \caption{The labyrinth tiling ${\cal L}_4$. The dashed lines 
represent the product ${\cal C}_{4} \times {\cal C}_{4}$.\label{fig:Fig1}}
\end{figure}

The labyrinth tiling can be constructed from the Euclidean product
${\cal C}_{m}\times {\cal C}_{m}$ of two such chains. This product is
a rectangular grid, thus its vertices can be classified into {\em
even}\/ and {\em odd}\/ vertices if they can be reached from the
origin by an {\em even}\/ or {\em odd}\/ number of steps along the
bonds, respectively. This is completely analogous to the even and the
odd sublattice of the square lattice. Connecting the {\em even}\/
vertices by diagonal bonds, we obtain a finite approximant ${\cal
L}_{m}$ of the labyrinth tiling ${\cal L}$. The {\em odd}\/ vertices,
when connected by diagonal bonds, form another labyrinth tiling ${\cal
L}_{m}^{\ast}$ that is dual to ${\cal L}_{m}$. We note that, due to
the palindromicity of $w_{m}$, ${\cal L}_{m}$ and ${\cal
L}_{m}^{\ast}$ just differ by a $90$ degree rotation.\cite{BGB} The
finite labyrinth tiling ${\cal L}_m$ consists of $N_{m}^{2}/2$
atoms. An example is shown in Fig.~\ref{fig:Fig1}. By construction,
the labyrinth tiling is symmetric with respect to reflection at the
main diagonal. Taking this diagonal as the $x$ axis and the direction
orthogonal to it as the $y$ axis, the coordinates of the vertices of
the labyrinth tiling, labeled by $k,l\in \mathbb{Z}$, can be written
as\cite{CC}
\begin{mathletters}
\begin{eqnarray}
x_{k,l} & = & u_{k}+u_{l}\\
y_{k,l} & = & u_{k}-u_{l}
\end{eqnarray}
\end{mathletters}%
where the coordinates with {\em even}\/ values of $k+l$ belong to
${\cal L}$, those with {\em odd}\/ values of $k+l$ to ${\cal
L}^{\ast}$. Here,
\begin{equation}
u_{k} = k/\sqrt{2}+\left[k/\sqrt{2}\right]
\end{equation}
where $\left[x\right]$ denotes the integer closest to $x$. It is easy
to see that the sequence of long and short lengths given by $u_{k}$
again follows the octonacci sequence, but now the two intervals have
lengths $(\lambda\pm1)/2$ which again have ratio
$(\lambda+1)/(\lambda-1)=\lambda$. Thus, the diagonal of the labyrinth
$x_{k,k}=2u_{k}$ is just a $\sqrt{2}$-scaled version of the original
octonacci sequence.

\section{Tight-binding model}
\label{sec2b}

The energy spectra for tight-binding Hamiltonians on the labyrinth
tiling were investigated by Sire.\cite{CC} For properly chosen
Hamiltonians, the analysis reduces to the one-dimensional case, and
the energy spectrum can be derived directly from those of the
corresponding Hamiltonian on the octonacci chain. However, the
properties of eigenstates, which also factorize into the product of
two eigenstates of the octonacci chain, were not discussed in
Ref.~\onlinecite{CC}. Here, we follow the same route to study the
eigenvalues and eigenstates.

Consider two identical octonacci chains in the framework of a
tight-binding model with zero on-site potentials
\begin{mathletters}
\begin{eqnarray}
H^{(1)}\psi_{k}^{(1,i)} & = & 
t_{k}^{}\psi_{k-1}^{(1,i)}+t_{k+1}^{}\psi_{k+1}^{(1,i)} =
E^{(1,i)}\psi_{k}^{(1,i)}\label{eq:toc1}\\
H^{(2)}\psi_{l}^{(2,j)} & = & 
t_{l}^{}\psi_{l-1}^{(2,j)}+t_{l+1}^{}\psi_{l+1}^{(2,j)} =
E^{(2,j)}_{}\psi_{l}^{(2,j)}
\label{eq:toc2}
\end{eqnarray}
\end{mathletters}%
where superscripts $(1)$ and $(2)$ label the two chains and the
indices $i$ and $j$ enumerate the eigenfunctions $\psi$ and
eigenvalues $E$ of the two octonacci chains. Throughout the paper, we
employ free boundary conditions, which formally corresponds to setting
$\psi_{0}=\psi_{N_{m}+1}=0$.  The hopping parameters $t_{k}$ and
$t_{l}$ take values according to the octonacci sequence. We associate
$t_{k,l}=1$ to a {\em long}\/ bond of length $\lambda$ and $t_{k,l}=v$
to a {\em short}\/ bond of length $1$, respectively. 

The eigenvalues of the octonacci chain are symmetric with respect to
$E=0$; if $\psi$ is an eigenstate of $H$ with eigenvalue $E$, then the
state $\tilde{\psi}$ with amplitudes
\begin{equation}
\tilde{\psi}_{k} = (-1)^{k}\psi_{k}
\label{eq:tilde}
\end{equation}
is again an eigenstate, but has an eigenvalue $-E$. For $E=0$, the eigenvalue
equation reduces to the recursion
\begin{equation}
\psi_{k+1} = -\frac{t_{k}}{t_{k+1}}\psi_{k-1}
\end{equation}
which always yields precisely {\em two}\/ linearly independent
solutions $\psi^{\pm}$ which can be chosen to vanish on either {\em
even}\/ or {\em odd}\/ sites. These have the form
\begin{equation}
\psi^{-}_{2r-1} = (-1)^{r-1}\psi^{-}_{1}
\prod_{s=2}^{r}\frac{t_{2s-2}}{t_{2s-1}},\qquad
\psi^{-}_{2r} = 0,
\label{eq:psim}
\end{equation}
and
\begin{equation}
\psi^{+}_{2r-1} = 0, \qquad
\psi^{+}_{2r} = (-1)^{r-1}\psi^{+}_{2}
\prod_{s=2}^{r}\frac{t_{2s-1}}{t_{2s}},
\label{eq:psip}
\end{equation}
where $\psi^{-}_{1}\neq 0$ and $\psi^{+}_{2}\neq 0$ are determined, up
to phases, by normalization. We note that one has to be careful if one
employs periodic boundary conditions because these, for an odd length
of the chain, couple the even and odd sublattices of the chain.  Thus,
while there are again two states (\ref{eq:psim}) and (\ref{eq:psip})
for a periodic chain of even length, only a single state at $E=0$
exists for an odd length of the chain.

Multiplying the two Eqs.~(\ref{eq:toc1}) and (\ref{eq:toc2}), we
obtain
\begin{eqnarray}
H^{(1,2)}\Phi_{k,l}^{(i,j)} & = & 
t_{k}^{}t_{l}^{}\Phi_{k-1,l-1}^{(i,j)}+
t_{k}^{}t_{l+1}^{}\Phi_{k-1,l+1}^{(i,j)}+
t_{k+1}^{}t_{l}^{}\Phi_{k+1,l-1}^{(i,j)}+
t_{k+1}^{}t_{l+1}^{}\Phi_{k+1,l+1}^{(i,j)}\nonumber\\ 
& = & E^{(1,i)}_{}E^{(2,j)}_{}\Phi_{k,l}^{(i,j)}
\label{eq:prod}
\end{eqnarray}
where we defined
\begin{equation}
\Phi_{k,l}^{(i,j)} =  \psi_{k}^{(1,i)}\psi_{l}^{(2,j)}
\label{eq:prodef}
\end{equation}
as an eigenfunction on the product of the two chains with eigenvalue
$E^{(1,i)}_{}E^{(2,j)}_{}$.  In Eq.~(\ref{eq:prod}), only wave
function amplitudes at positions $(k\pm 1,l\pm 1)$ contribute, thus
the Hamiltonian $H^{(1,2)}$ corresponds to hopping along the {\em
diagonals}\/ of the product grid ${\cal C}_{m}\times {\cal C}_{m}$. The
corresponding hopping parameters are the product of two hopping
parameters in the octonacci chain and thus take values $1$, $v$, and
$v^2$ for diagonals of length $\lambda+1$, $\sqrt{2\lambda+2}$, and
$\lambda-1$, respectively. 

Thus the system in Eq.~(\ref{eq:prod}) naturally separates into {\em
two}\/ independent sets of equations with $k+l$ {\em even} or $k+l$
{\em odd}, respectively. In this paper, we restrict our investigation
to the case with $k+l$ even as the other case is completely analogous.
Thus, $H^{(1,2)}$ can be interpreted as a tight-binding Hamiltonian
with zero on-site potential defined on the labyrinth tiling ${\cal
L}$. Clearly, the eigenvalues for the labyrinth are just products of
the eigenvalues for the octonacci chain, and all such products appear
as eigenvalues because the spectra of the two dual labyrinth tilings
${\cal L}_{m}$ and ${\cal L}_{m}^{\ast}$ are identical. For the
corresponding eigenfunctions on ${\cal L}$, we have to construct
linear combinations of the product eigenfunctions $\Phi_{i,j}$
(\ref{eq:prodef}) which vanish on the vertices of the dual tiling
${\cal L}_{m}^{\ast}$. This can be done as follows.

Suppose $\psi^{(1,i)}$ and $\psi^{(2,j)}$ are normalized
eigenfunctions of the octonacci chain with eigenvalues $E^{(1,i)}$ and
$E^{(2,j)}$, respectively. Then both 
$\Phi^{(i,j)}=\psi^{(1,i)}\psi^{(2,j)}$ and
$\tilde{\Phi}^{(i,j)}=\tilde{\psi}^{(1,i)}\tilde{\psi}^{(2,j)}$
(\ref{eq:tilde}) are eigenfunctions of $H^{(1,2)}$ with the same
eigenvalue $E^{(1,i)}E^{(2,j)}$, where we assume $E^{(1,i)}\neq 0$ and
$E^{(2,j)}\neq 0$. But from Eq.~(\ref{eq:tilde}) we have
\begin{equation}
\tilde{\Phi}^{(i,j)}_{k,l} = (-1)^{k+l}\Phi^{(i,j)}_{k,l}
\end{equation}
and thus the linear combinations
\begin{equation}
{\Psi^{(i,j)}}^{\pm} =
\frac{1}{\sqrt{2}}\left(\Phi^{(i,j)}\pm\tilde{\Phi}^{(i,j)}\right)
\label{eq:lc1}
\end{equation}
are normalized eigenfunctions that vanish for {\em odd}\/ or for {\em
even}\/ values of $k+l$, and thus on ${\cal L}^{\ast}$ or ${\cal L}$,
respectively. If one or both eigenvalues $E^{(1,i)}$ and $E^{(2,j)}$
are zero, we can make use of the previously discussed eigenfunctions
$\psi^{+}$ (\ref{eq:psip}) and $\psi^{-}$ (\ref{eq:psim}) to construct
the desired wavefunctions. If one of the eigenvalues vanishes, say,
without loss of generality, $E^{(1,i)}=0$ and $E^{(2,j)}\neq 0$, the
appropriate four linear combinations are
\begin{equation}
{\Psi^{(i,j)}}^{\pm\pm} =
\frac{1}{\sqrt{2}}{\psi^{(1)}}^{\pm}
\left(\psi^{(2,j)}\pm\tilde{\psi}^{(2,j)}\right)
\label{eq:lc2}
\end{equation}
where ${\psi^{(1)}}^{\pm}$ is the wavefunction of Eqs.~(\ref{eq:psip})
and (\ref{eq:psim}) on the first chain, and we also used the state
$\tilde{\psi}^{(2,j)}$ which has an energy $-E^{(2,j)}$. Clearly, the
wave functions ${\Psi^{(i,j)}}^{++}$ and ${\Psi^{(i,j)}}^{--}$ have
support on ${\cal L}$, the other two linear combinations
${\Psi^{(i,j)}}^{+-}$ and ${\Psi^{(i,j)}}^{-+}$ live on ${\cal
L}^{\ast}$.  Finally, for $E^{(1,i)}=E^{(2,j)}=0$, we have four states
\begin{equation}
{\Psi^{(i,j)}}^{\pm\pm} =
{\psi^{(1)}}^{\pm}{\psi^{(2)}}^{\pm}
\label{eq:lc3}
\end{equation}
where again ${\Psi^{(i,j)}}^{++}$ and ${\Psi^{(i,j)}}^{--}$ are
supported on ${\cal L}$, and the remaining two product states
${\Psi^{(i,j)}}^{+-}$ and ${\Psi^{(i,j)}}^{-+}$ on ${\cal L}^{\ast}$.
In particular, this argument proves that $E=0$ is a $(2N_{m}\! -\!
2)$-fold degenerate eigenvalue for the labyrinth ${\cal L}_m$. Thus we
find, as for simple tight-binding Hamiltonians on the
Penrose\cite{confined} or the octagonal Ammann-Beenker tiling, a large
degeneracy of states in the ``band'' center at $E=0$.  However, in
contrast to these well-known examples where the degeneracy stems from
certain ``confined'' states\cite{confined} that occur as a
consequence of the local topology of the tilings, the spectral measure
carried by the states at $E=0$ vanishes for the labyrinth as
$N_{m}\rightarrow\infty$, thus it is not a finite fraction of the
eigenstates that contributes to $E=0$ in this case.

In practice, having the complete knowledge of the eigenstates for the
labyrinth tiling at our disposal, we do not need to care too much
about the precise linear combinations of states derived above. Since
the eigenvalues $E_{i}$, $i=1,\ldots,N$ of the octonacci chain are
symmetric about zero, one can obtain the set of eigenvalues of the
labyrinth tiling simply as
\begin{equation}
\left\{E_{i}E_{j} \mid 1\leq i\leq {\textstyle\frac{N}{2},\; j\leq i}\right\} \cup
\left\{E_{i}E_{j} \mid {\textstyle\frac{N}{2}}<i\leq N,\; j\leq i-1\right\}
\end{equation}
where we assume that the eigenvalues of the octonacci chain are
ordered as $E_{i}\geq E_{j}$ for $i>j$. The corresponding eigenvectors
are most easily constructed by just restricting the products of
eigenvectors to the sites of the labyrinth ${\cal L}$, and
re-normalizing the resulting eigenstate. Eq.~(\ref{eq:tilde})
guarantees that this procedure yields the correct results, because the
states $\psi$ and $\tilde{\psi}$ just differ by an alternating sign.

\section{Energy spectra and wavefunctions}
\label{sec3}

\begin{figure}
 \centerline{\psfig{figure=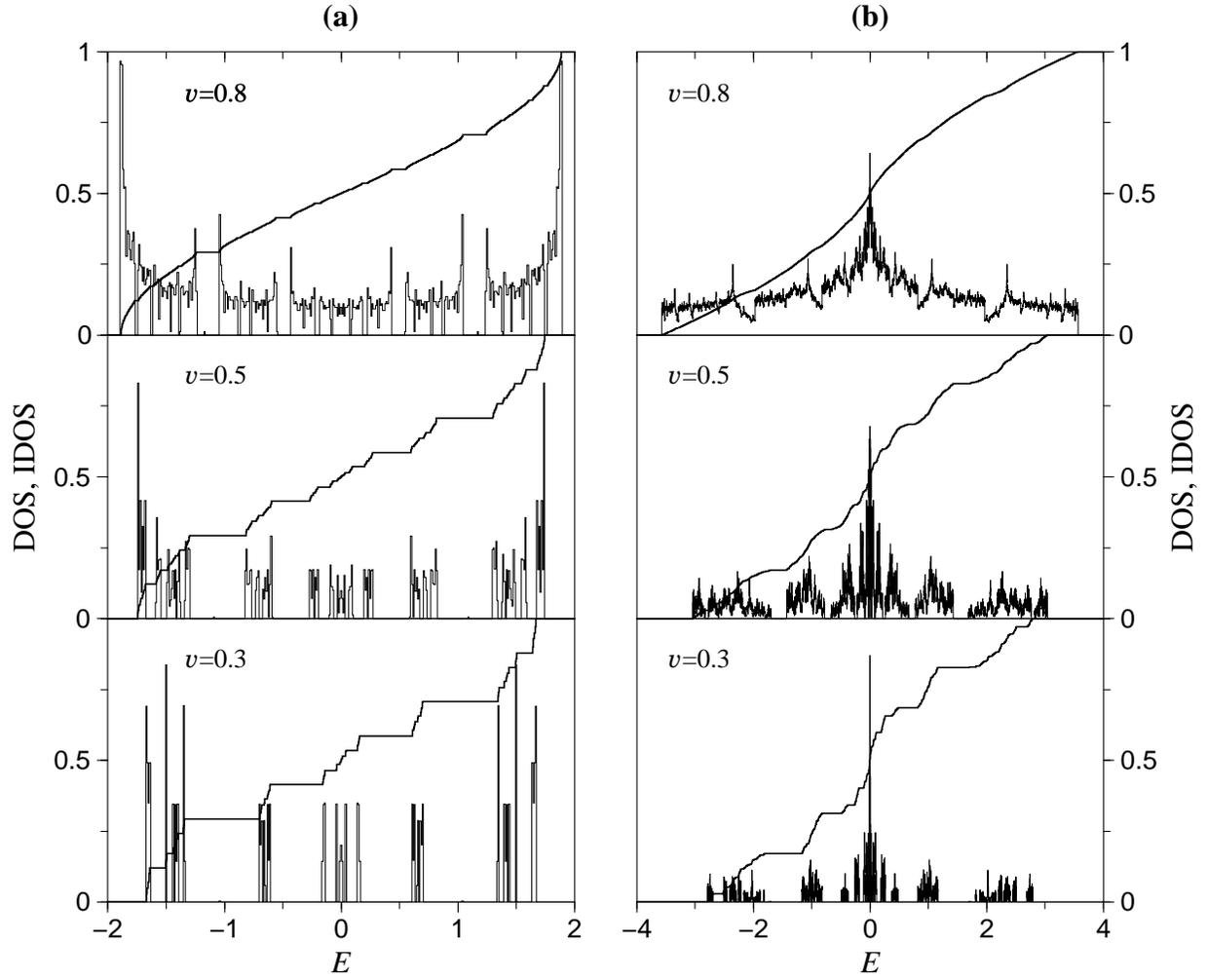,width=\columnwidth}}
 \caption{The DOS (arbitrary scale) and the IDOS (bold line) for 
          various hopping parameters $v$ for (a) 
          the octonacci chain, and (b) the labyrinth tiling.\label{fig:Fig2}}
\end{figure}

Following the results of Sec.~\ref{sec2b}, one can easily calculate
the density of states (DOS) and the integrated density of states
(IDOS). For comparison, we show the DOS and the IDOS for the octonacci
chain and the labyrinth tiling in Fig.~\ref{fig:Fig2}. For the
octonacci chain, the IDOS is a devil's staircase even for $v$ close to
$1$ and the DOS is singular continuous with zero Lebesgue measure. By
more detailed analysis, one finds a self-similar energy spectrum for
the octonacci chain with a hierarchical gap structure as described by
the gap labelling theorem.\cite{gaplabel} In contrast, we observed a
smooth IDOS without visual gaps as $v$ approaches $1$ in the labyrinth
tiling. A more detailed analysis of the IDOS and the energy spectra
shows that in the regime $0.6<v<1.0$ the energy spectrum contains no
or a finite number of gaps; for $v<0.6$ the spectrum is fractal-like
and the IDOS is similar to a devil's staircase. Sire found that the
spectrum is singular continuous with {\em finite}\/ Lebesgue measure
for $v\geq 0.4$,\cite{CC} which may indicate that the spectrum is a
mixture of band-like and fractal-like parts in the regime $0.4\leq
v<0.6$. In Fig.~\ref{fig:Fig2}(b) one can see a peak at the center of
the spectrum which is due to the degenerate states at $E=0$.  But it
differs from the localized states observed in the Penrose
tiling\cite{confined} in the sense that no leap at $E=0$ is seen in
the IDOS, in agreement with the results of the previous section. For
varying parameter $v$, we find three regions with different behaviour
of the DOS of the labyrinth tiling: a maximum around the center,
distinct shoulders located between the spectral center and edge, and a
tail at the band edge, which is similar to the behaviour observed for
a tight-binding model on the icosahedral Ammann-Kramer
tiling.\cite{TM}

In order to characterize the eigenstates, we employ a multifractal
analysis, which is based on the standard box-counting
procedure.\cite{TM,ARM} In our numerical calculations, we determine
the singularity strength $\alpha(q)$ and the corresponding fractal
dimension $f(q)$ by a linear regression procedure, but prior to this
we need to check the linearity of $\sum_{i}\mu_{i}\ln\mu_{i}$ versus
$\ln{\varepsilon}$, where $\mu_{i}(q,\varepsilon)$ denotes the
normalized $q$th moment of the box probability for boxes of linear
size $\varepsilon L$.  A homogenously extended wave function
corresponds to $\alpha(q)=f(q)=d$, where $d$ denotes the spatial
dimension. For critical eigenstates, the fractal dimension $f$ is a
smooth convex function of $\alpha$, and $\alpha$ should be limited to
a $q$-interval.  Moreover, the generalized dimensions of the
eigenstate $\psi$ are given by $D_{q}^{\psi}=\left[f(q)-q\alpha
(q)\right]/(1-q)$ for $q\neq 1$ and $D_{1}^{\psi}=f(1)=\alpha(1)$.

\begin{figure}
\centerline{\psfig{figure=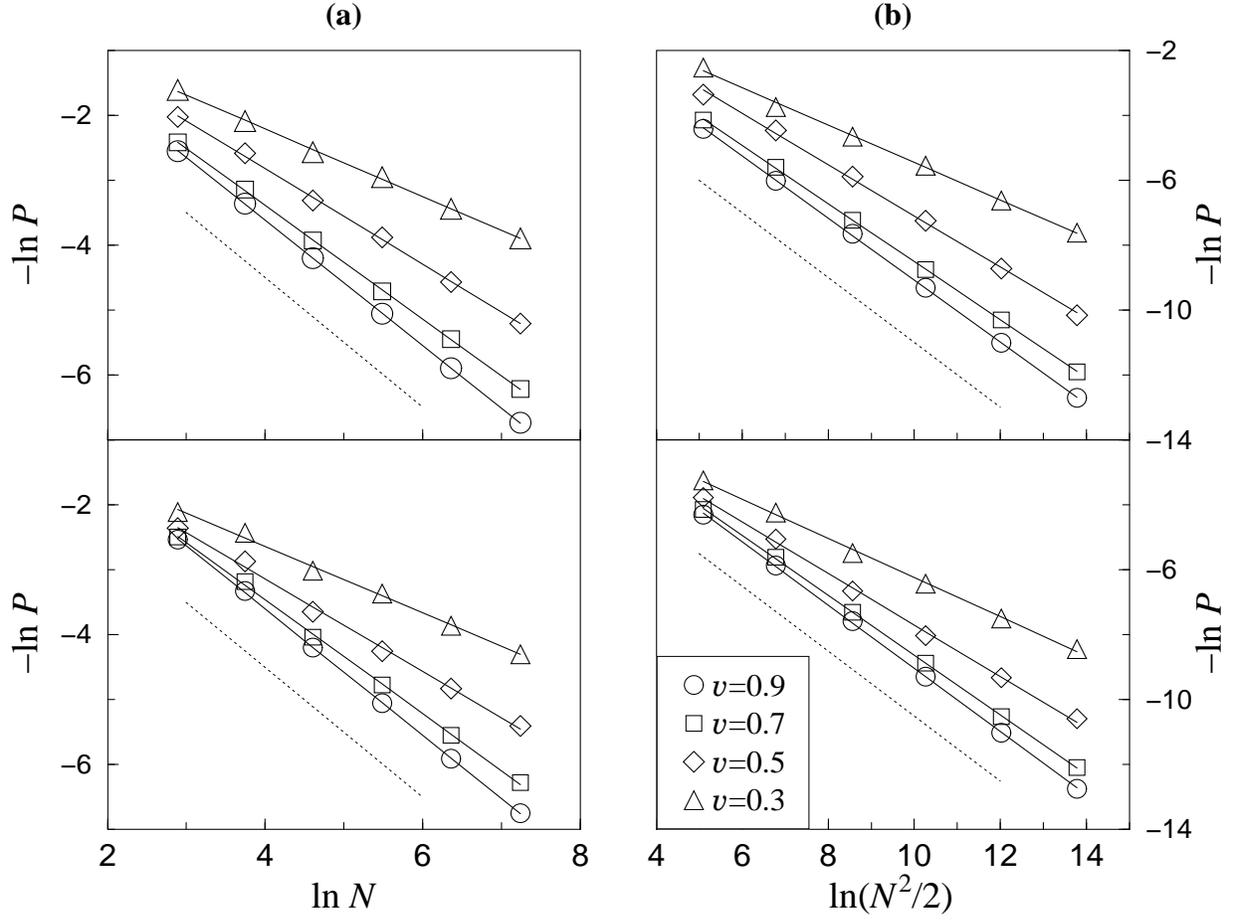,width=\columnwidth}}
\caption{The dependence of the inverse participation numbers 
$P^{-1}(E,V)$ on the size $V$ of (a) the octonacci chain, 
and (b) the labyrinth tiling,
for $E=0$ (top) and $E=E_{\text{min}}$ (bottom). 
Lines are least-squares fits to the data, the dotted lines correspond to
$P^{-1}(E,V)\sim V^{-1}$.
\label{fig:Fig4}}
\end{figure}

The singularity spectra $f(\alpha)$ of eigenstates for both the
octonacci chain and the labyrinth tiling show the typical shape
expected for multifractal states, thus we refrain from showing these
here.  For the octonacci chain, the eigenstates in the ``band'' center
are more extended than those at the ``band'' edge. In this case, the
curves $f(\alpha)$ become fairly narrow as $v$ approaches
$1$. Generally, the eigenstates show stronger multifractal
characteristics with decreasing parameter $v$. In contrast to the
behaviour observed for the Penrose tiling,\cite{TM} for the labyrinth
tiling we do not find that the multifractal behaviour of eigenstates
becomes significantly stronger when moving from energies at the edge
towards the center of the ``band''. We also calculated the scaling
behaviour of the inverse participation number
\begin{equation}
P^{-1}(E,V)=\sum_{\bf r}|\psi({\bf r})|^{4}
\label{eq:ipn}
\end{equation}
with respect to the size $V=L^{d}$ of the system,\cite{TM,ARM} i.e.,
\begin{equation}
P^{-1}(E,V) \sim V^{-\gamma(E)}
\label{eq:ipnscal}
\end{equation}
for large $V$. A fractal eigenstate is characterized by $0<\gamma<1$,
whereas $\gamma=0$ corresponds to a localized state, and $\gamma=1$ to
an extended state. In general, the scaling exponent $\gamma(E)$
depends on the energy.\cite{ARM} Numerically, one analyzes the scaling
behaviour of $P^{-1}(E,V)$ at an energy $E$ by averaging over the
eigenstates within a small energy interval $E\pm\Delta E/2$. The
result for eigenvectors from the center and at the lower edge of the
spectrum is shown in Fig.~\ref{fig:Fig4} which corroborates the
multifractal nature of the eigenstates in both systems. The exponent
$\gamma$, given by the slope, decreases, presumably continuously, from
$\gamma=1$ for the periodic case $v=1$ to $\gamma=0$ for $v=0$.

\section{Quantum diffusion for the octonacci chain}
\label{sec4}

In this short section, we briefly present our numerical results of the
autocorrelation function $C(t)$ and the mean square displacement
$d(t)$ for the octonacci chain. Further discussion and comparison with
the results for the labyrinth will be given below.

Fig.~\ref{fig:Fig5} shows the autocorrelation function $C(t)$ of the
octonacci chain. The initial site is located at the center of the
system. The long-time behaviour of $C(t)$ follows $C(t)\sim
t^{-\delta}$ with $0<\delta<1$ for different $v$. For small $v$,
$C(t)$ displays strong oscillatory behaviour, which may result from
level fluctuations.  The result for $d(t)$ is displayed in
Fig.~\ref{fig:Fig6}. Evidently, $d(t)\sim t^{\beta}$ and $\beta$
increases with increasing $v$, limited by $\beta<1$. For a given
modulation parameter $v$, we observe the relation $\beta>\delta$
between the two exponents. Similar results have been obtained for 1D
Fibonacci chains and at the mobility edge of the Harper
model.\cite{RGT,JR,SD,HHTM} Therefore, in accordance with the singular
continuous energy spectra and the multifractal eigenstates, the
diffusion is usually anomalous in 1D quasiperiodic systems.

\section{Quantum diffusion for the labyrinth tiling}
\label{sec5}

We now switch to the more interesting case of the labyrinth tiling. In
Fig.~\ref{fig:Fig7}, we show the behaviour of $C(t)$ for the labyrinth
tiling. The number of sites in our system is $N^2/2=19\,
602^{2}/2=192\, 119\, 202$, which is much larger than other 2D
quasiperiodic systems discussed previously such as, for instance,
Fibonacci lattices\cite{JR} and the octagonal tiling.\cite{HJ}
Therefore, we can utilize this system to study the long-time behaviour
of $C(t)$ more accurately than before. Apparently, Fig.~\ref{fig:Fig7}
again exhibits a power law behaviour $C(t)\sim t^{-\delta}$.  By a
more detailed analysis, we surprisingly find a transition point at
$v_{c}\approx 0.6$. For $v<v_c$ the slope of the curves decreases with
decreasing $v$. In the regime $v>v_c$, the behaviour of $C(t)$ is the
same as for a periodic system, i.e., $C(t)\sim t^{-1}$.\cite{JR} When
compared to the results of Sec.~\ref{sec3}, we see that this regime
corresponds to the region where one finds band-like energy
spectra. Since $\delta$ equals the correlation dimension $D_{2}$ of
the energy spectral measure,\cite{RGT,JR} $\delta=1$ is reasonable for
the case of band-like spectra. Similar to the 1D case, one still has
$0<\delta<1$ in the regime $v<v_c$ with fractal-like or mixed spectra
for the labyrinth tiling. We expect that this is a general result for
higher-dimensional quasiperiodic systems.  Furthermore, we find the
behaviour of $C(t)$ is independent of the initial site, which can be
observed from the example shown in Fig.~\ref{fig:Fig7}. Of course, as
our analysis is based on numerical data for a finite system, we cannot
possibly {\em prove}\/ the existence of a true transition point
$v_{c}$, because we cannot rule out a rapid, but continuous change in
$\delta$ around $v\approx 0.6$.

The calculation of the square displacement $d(t)$ is numerically more
expensive, thus we restrict ourselves to a smaller system of
$N^2/2=578^{2}/2=167\, 042$ sites. Nevertheless, this is still larger
than the 2D octagonal quasicrystals studied previously.\cite{BCV} In
Fig.~\ref{fig:Fig8}, we show that the long-time behaviour is described
by a power law $d(t)\sim t^{\beta}$. In contrast to $C(t)$, we do not
find a transition point for $d(t)$ as the parameter $v$ is varied.  As
for the octagonal tiling\cite{BCV} and for the octonacci chain,
$0<\beta<1$ for the labyrinth tiling. Therefore, a band spectrum does
not imply ballistic diffusion in quasicrystals. It can be argued that
the exponent $\beta$ is associated with the correlation dimension
$D_{2}^\psi$ of the eigenstates.\cite{RKST} In 1D quasiperiodic
systems, or at the metal-insulator transition in the Anderson model of
localization, the eigenstates are multifractal\cite{1D,MHT,ARM} and
$0<\beta<1$.\cite{SD,HHTM,TT} In accordance, the multifractal
eigenstates in 2D quasicrystals may be expected to lead to anomalous
diffusion with $0<\beta<1$. Possibly, ballistic diffusion can occur in
3D quasicrystals because their wavefunctions are more
extended.\cite{TM}

So far, we assumed that the initial wave packet is a
$\delta$-function, thus we start with an electron that is localized at
a particular site $n_{0}=(k_{0},l_{0})$ and follow the spreading of
its wave function $\Psi^{\{n_{0}\}}$ with time. This means that, in
general, {\em all}\/ eigenstates contribute to the time evolution
because the expansion in terms of the orthonormal basis of eigenstates
$\Psi^{(i,j)}$  is
\begin{equation}
\Psi^{\{n_{0}\}}_{k,l} = 
\delta_{k,k_{0}}^{}\delta_{l,l_{0}}^{} = 
\sum_{i,j}\Psi^{(i,j)}_{k_{0},l_{0}}\Psi^{(i,j)}_{k,l}
\label{eq:delta}
\end{equation}
and thus the entire energy spectrum is probed. For convenience, we
dropped the superscripts $\pm$ on the wavefunctions of
Eqs.~(\ref{eq:lc1})--(\ref{eq:lc3}), assuming that the proper linear
combinations are used that are supported on the labyrinth ${\cal
L}_{m}$. Note that we do not need a complex conjugation in
Eq.~(\ref{eq:delta}) because the Hamiltonian is a real symmetric and
we therefore can choose eigenvectors that form a real orthogonal
matrix. In order to check for an energy dependence of the diffusion,
we now consider different initial wave packets
$\Psi^{\{n_{0},[E-\frac{\Delta E}{2},E+\frac{\Delta E}{2}]\}}$ which
have a finite width and are constructed as linear combinations of
eigenstates from a certain energy window $[E-\Delta E/2,E+\Delta
E/2]$. The new normalized states can be written as
\begin{equation}
\Psi^{\{n_{0},[E-\frac{\Delta E}{2},E+\frac{\Delta E}{2}]\}}_{k,l} = 
\frac{\sum^{\prime}_{i,j}\Psi^{(i,j)}_{k_{0},l_{0}}\Psi^{(i,j)}_{k,l}}
{\sqrt{\sum^{\prime}_{i,j}|\Psi^{(i,j)}_{k_{0},l_{0}}|^{2}}}
\label{eq:wavepack}
\end{equation}  
where the sum $\sum^{\prime}$ is restricted to the eigenstates
$\Psi^{(i,j)}$ with eigenvalues $E^{(1,i)}E^{(2,j)}\in [E-\Delta
E/2,E+\Delta E/2]$.\cite{Schw} Clearly, Eq.~(\ref{eq:wavepack})
becomes Eq.~(\ref{eq:delta}) if the energy interval contains the
complete spectrum, this is nothing but the usual completeness
condition of the basis of eigenvectors.

We numerically checked different energy windows for the octonacci
chain and the labyrinth tiling. Due to the high DOS around $E=0$, we
choose smaller intervals in the band center. The results in
Fig.~\ref{fig:Fig9} and Fig.~\ref{fig:Fig10} show that the long-time
behaviour of $C(t)$ and $d(t)$ hardly depends on the selection of the
energy window.  However, it is more complex at small times due to the
different shapes and widths of the initial wave packets. The various
values of $d(t)$ at the initial time reflect the width of the initial
wave packet. The smaller the energy interval under consideration, the
wider is the initial wavepacket. In practice, in order to avoid that
the wave packet reaches the boundary too early, the energy interval
may not be chosen too small.

\section{Dynamical scaling and fractal dimensions}
\label{sec6}

In 1D quasiperiodic systems, it is known that the inequality
$\beta\geq D_{1}$ relates the diffusion behaviour and the fractal
properties of the energy spectrum.\cite{IG} In $d$ dimensions, this
generalizes to the inequality $\beta\geq D_{1}/d$, thus it implies a
superdiffusive behaviour $\beta\geq 1/2$ in two dimensions if $D_1\ge
1$.  In Fig.~\ref{fig:Fig11}, the values of the exponent $\beta$ for
the octonacci chain and the labyrinth tiling are shown for various
values of the parameter $v$. In all cases, we find that this
inequality holds. Apparently, the diffusion exponents $\beta$ for the
octonacci chain and the labyrinth tiling are very close, which might
be due to the product structure of the labyrinth and its
wavefunctions.

According to a conjecture by Pi\'{e}chon, \cite{FP} $\beta=D_{-1}$ of
the global spectral measure for one-dimensional quasiperiodic models
with multifractal global spectral measure. In order to check this
relation, we calculated $D_{-1}$, but it turns out that is rather
difficult to extract the accurate values by a linear fit due to strong
oscillations in the data. However, it appears that the relation does
not hold for general parameter values $v$ in the octonacci chain, and
it certainly cannot be valid for the two-dimensional system as it only
involves a dimension that characterizes the spectral measure.

Ketzmerick {\it et al.}\cite{RKST}\/ suggested an improved inequality
$\beta \geq D_{2}/D_{2}^{\psi}$ which is numerically obeyed by 1D
quasiperiodic models.\cite{RKST} As can be seen from
Fig.~\ref{fig:Fig11}, this relation applies for the octonacci chain as
well as for the the labyrinth tiling. However, in the two-dimensional
case the inequality is less sharp as $\beta$ is much larger than the
ratio $D_{2}/D_{2}^{\psi}$, in particular for values of the parameter 
$v\ge 0.6$ where the energy spectrum is smooth and $D_{2}\approx 1$.

For multifractal wavefunctions at the Anderson transition or at
quantum Hall transitions, one finds $D_{2}^{\psi}=d D_{2}$ for a
$d$-dimensional system.\cite{BL,BR} Above, it has been demonstrated
that $D_{2}=1$ for the band-like spectra in 2D quasiperiodic systems,
but the corresponding eigenstates are multifractal with generalized
dimension $D_{2}^{\psi}<2$. Although the eigenstates of 2D
quasiperiodic tight-binding models are similar to the critical states
at the Anderson transition, the equality $D_{2}^{\psi}=d D_{2}$
apparently does not apply to 2D quasicrystals.

Recently, Zhong {\it et al.}\cite{zhong}\/ argued that one might
interpret the superdiffusive behaviour in aperiodic systems as a
ballistic behaviour in a space of effective dimension $D_{2}^{\psi}$,
or that this should at least give an upper bound on the possible
values of $\beta$. In Fig.~\ref{fig:Fig11}, we compare the ratio
$D_{2}^{\psi}/d$ to $\beta$. It turns out that the values of $\beta$
and $D_{2}^{\psi}/d$ are rather close, but that there seems to be a
systematic deviation with $\beta<D_{2}^{\psi}/d$ for small values of
$v$ and $\beta>D_{2}^{\psi}/d$ for $v$ close to $1$. Therefore, at
least for large values of the parameter $v$, it appears that this
bound does not hold.

Finally, we also included the values of $D_1^{\psi}/d$ in
Fig.~\ref{fig:Fig11}, which apparently does give an upper bound on the
values of $\beta$ for the models under consideration. So far, this is
just an observation, we cannot present an argument that this should
hold in general. 

\section{Conclusion}
\label{sec7}

In this paper, the energy spectra, wavefunctions and quantum diffusion
for the octonacci chain and the labyrinth tiling are studied. The
labyrinth tiling is based on the octonacci chain, which allows us to
deal with very large systems. For the octonacci chain, the energy
spectra are singular continuous and the eigenstates are critical. The
energy spectra of the labyrinth tiling presumably are also singular
continuous, but they can be band-like (i.e., of finite Lebesgue
measure) with zero or finite gaps, a mixture of band-like and fractal
parts, or fractal-like upon increasing the modulation
strength. However, the eigenstates are multifractal irrespective of
the value of the modulation parameter. 

The propagation of an initial wave packet is discussed in terms of the
autocorrelation function $C(t)$ and the mean square displacement
$d(t)$.  Numerical results show that $C(t)\sim t^{-\delta}$ and
$d(t)\sim t^{\beta}$ for the octonacci chain and the labyrinth
tiling. Corresponding to the multifractal eigenstates, we observe
$0<\beta<1$ for both systems. In the case of fractal-like or mixed
energy spectra and multifractal eigenstates, we find $0<\delta<1$.
However, for a band-like spectrum, $C(t)\sim t^{-1}$ as in a periodic
system, which causes a qualitative change of behaviour in $C(t)$ for
the labyrinth tiling at a parameter value $v_c\approx 0.6$.  Similar
effects have also been observed for Fibonacci lattices\cite{JR} and
for the octagonal tiling.\cite{HJ} 

We believe that the anomalous diffusion shown in $d(t)$ and the
crossover of the autocorrelation $C(t)$ will be a common phenomenon in
2D quasiperiodic systems. Of course, to observe the crossover in
$C(t)$ one needs a parameter that allows one to continuously move away
from the periodic case, which is not easily at hand for the most
commonly investigated quasiperiodic model systems such as the Penrose
or the octagonal tiling. Finally, we also studied the influence of
different initial wave packets by choosing the eigenstates from
various energy windows. The results show that the behaviour of $C(t)$
and $d(t)$ does not depend significantly on the shape and the location
of the initial wave packet.

Comparing the values of $\beta$ with several expressions involving the
fractal dimensions of energy spectra and eigenstates that were
proposed in the literature, we find that the inequality $\beta \geq
D_{2}/D_{2}^{\psi}$ of Ref.~\onlinecite{RKST} holds true. However, it
seems that the bound $\beta\le D_{2}^{\psi}/d$ proposed recently by
Zhong {\it et al.}\cite{zhong}\/ may be violated for parameter values
$v$ close to one, i.e., close to the periodic case. However, we find
that the weaker condition $\beta\le D_1^{\psi}/d$ is always satisfied.

Our present work corroborates that there are strong relations between
fractal properties of energy spectra and wavefunctions on the one hand
and the exponents describing the quantum diffusion on the other hand.
However, it appears to be difficult to find relations that give
quantitative agreement for one- and two-dimensional aperiodic systems.
Here, a deeper understanding of the underlying physics is desirable.
Higher-dimensional systems constructed as products of one-dimensional
systems, such as the labyrinth tiling, provide useful toy examples for
further investigations which can, at least, be treated numerically in
an efficient way.

\acknowledgements 
The authors thank J.\ X.\ Zhong for fruitful discussions. HQY is
grateful for the kind hospitality in Chemnitz. Financial support from
DFG (UG) and the NSF of China (HQY) is gratefully acknowledged.

\clearpage

\clearpage

\clearpage

\clearpage
\begin{figure} 
\centerline{\psfig{figure=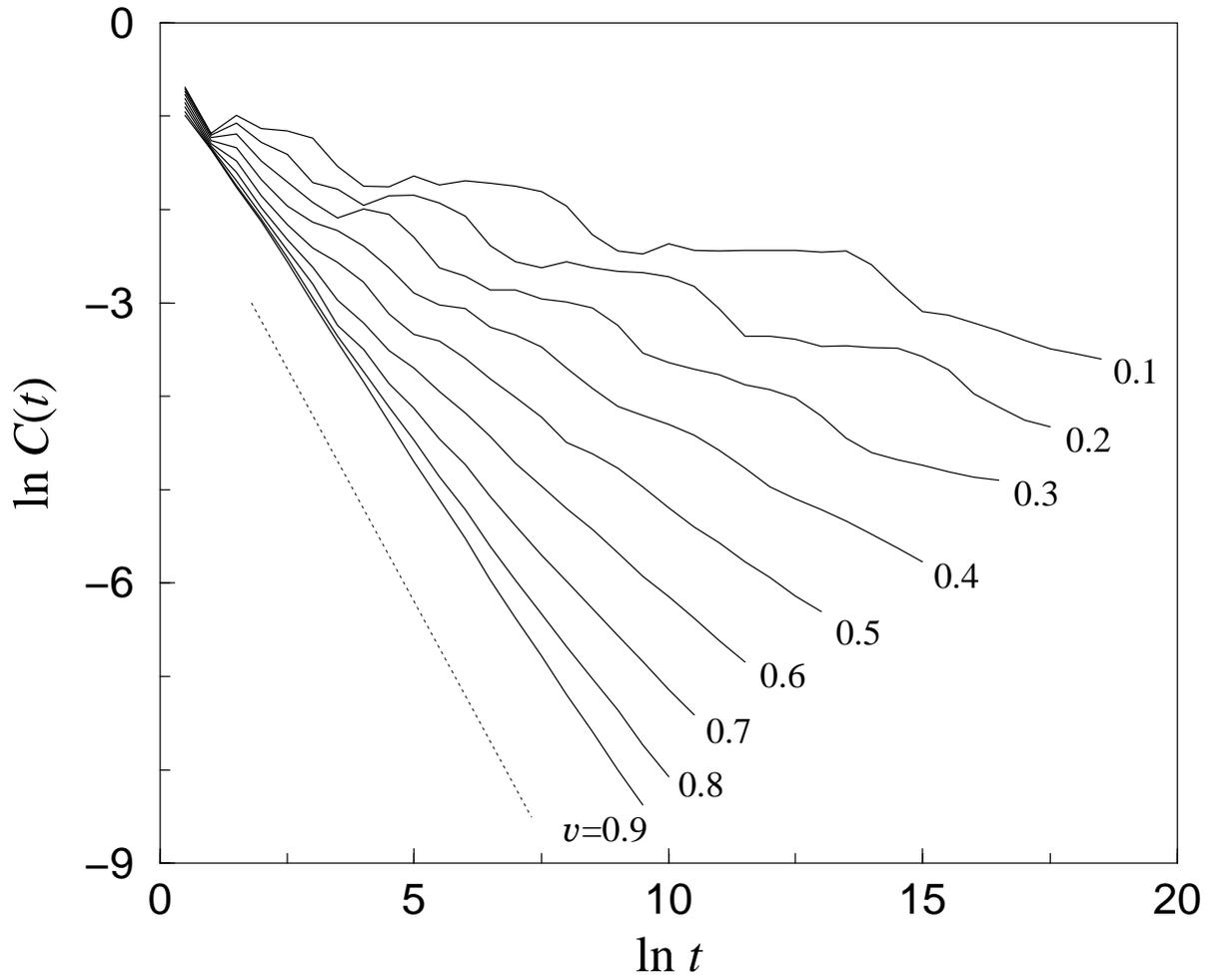,width=\columnwidth}}
 \caption{The autocorrelation function $C(t)$ for 
the octonacci chain with $N=19\, 602$ and initial site $n_{0}=9801$.
The dotted line corresponds to $C(t)\sim t^{-1}$.\label{fig:Fig5}}
\end{figure}

\clearpage
\begin{figure} 
\centerline{\psfig{figure=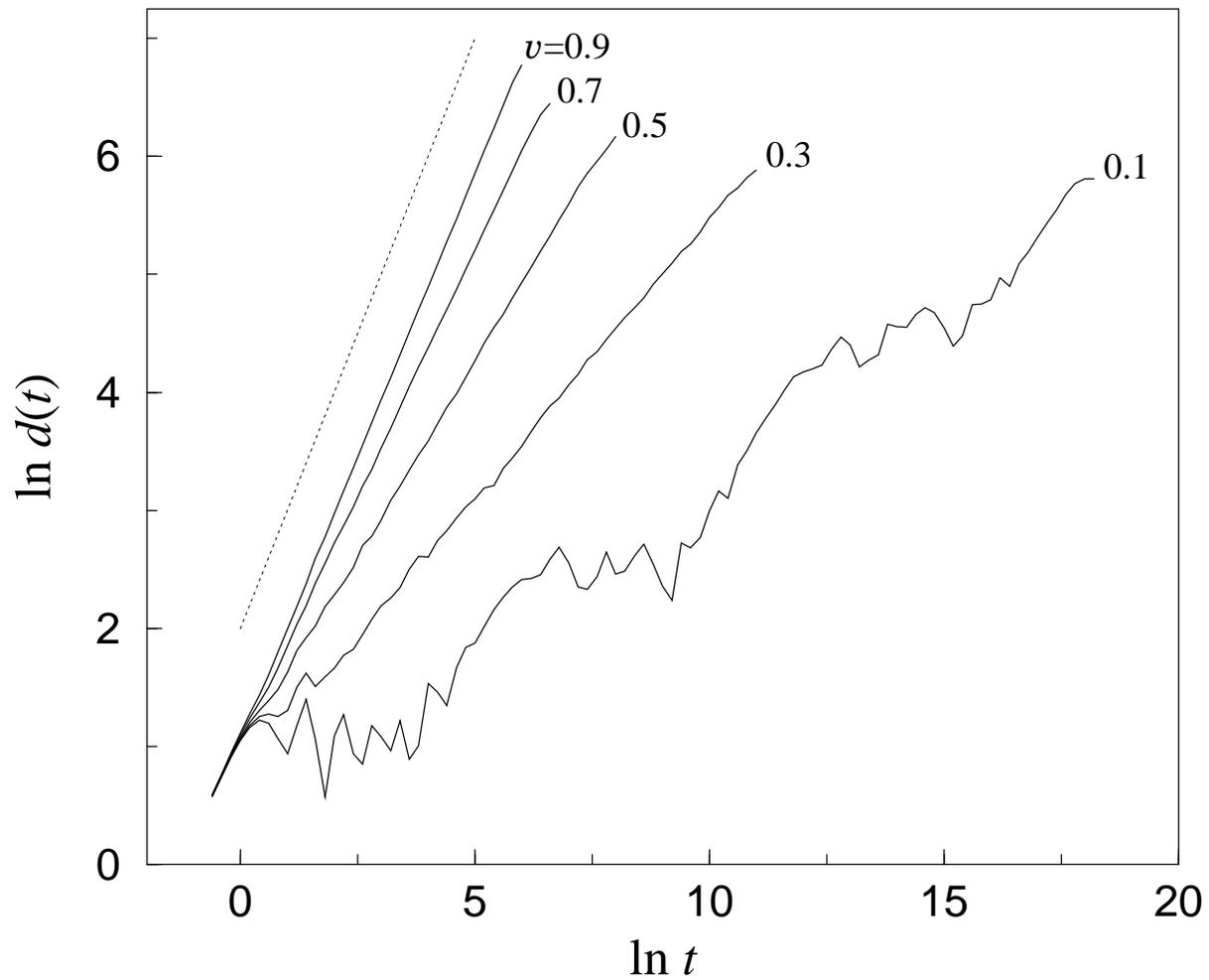,width=\columnwidth}}
 \caption{The mean square displacement $d(t)$ for the octonacci chain with
$N=1394$ and initial site $n_{0}=697$.
The dotted line corresponds to ballistic motion $d(t)\sim t$.
\label{fig:Fig6}}
\end{figure}

\clearpage
\begin{figure} 
\centerline{\psfig{figure=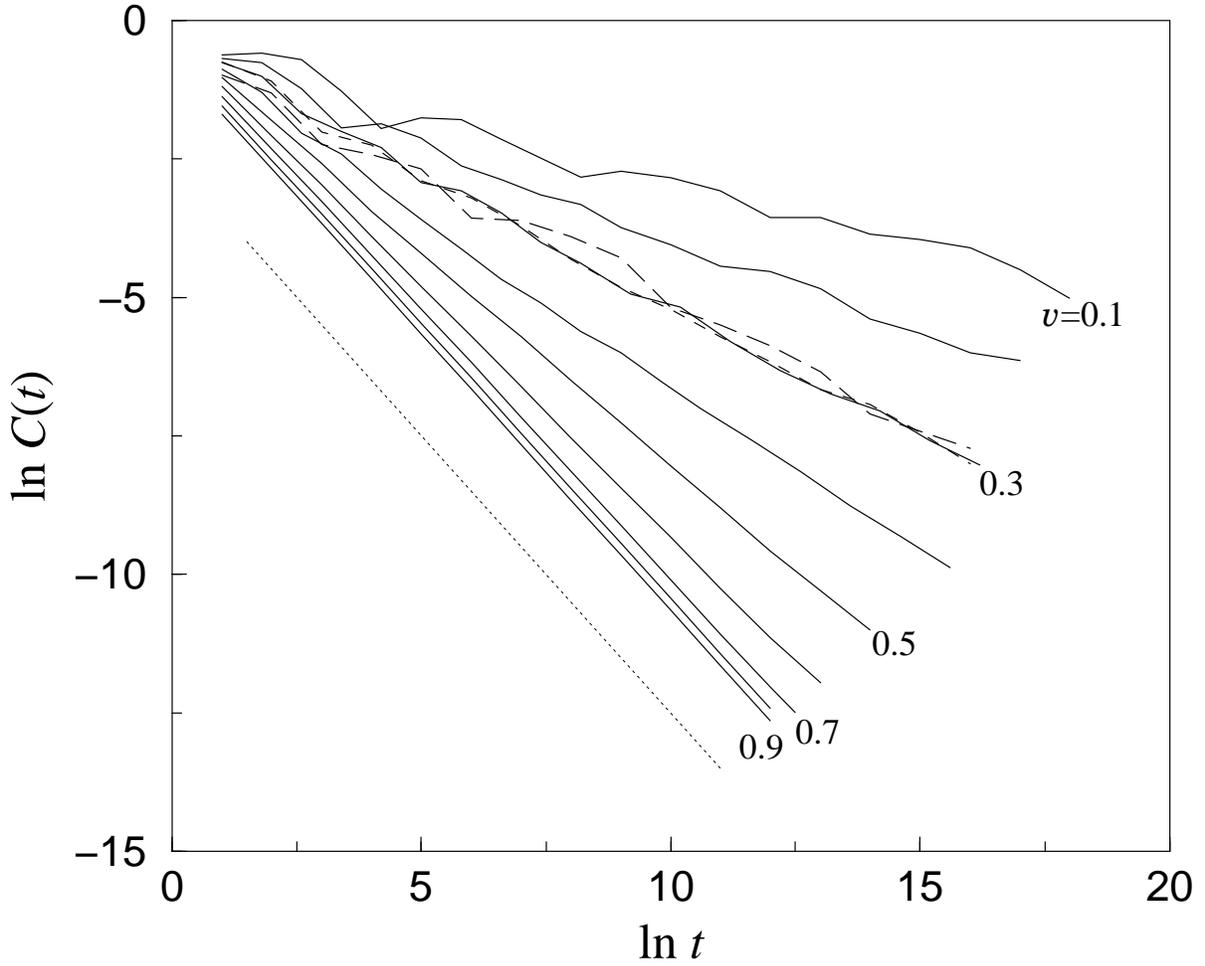,width=\columnwidth}}
 \caption{The autocorrelation function $C(t)$ for the labyrinth tiling with 
$v=0.1,0.2,\ldots,0.9$. The system size is
$N^2/2=19\, 602^{2}/2=192\, 119\, 202$,
and the initial site is $n_{0}=(9800,9800)$. For $v=0.3$, 
the two dashed lines correspond
to different choices of the initial site $n_{0}$.
The dotted line shows $C(t)\sim t^{-1}$.\label{fig:Fig7}} 
\end{figure}

\clearpage
\begin{figure} 
\centerline{\psfig{figure=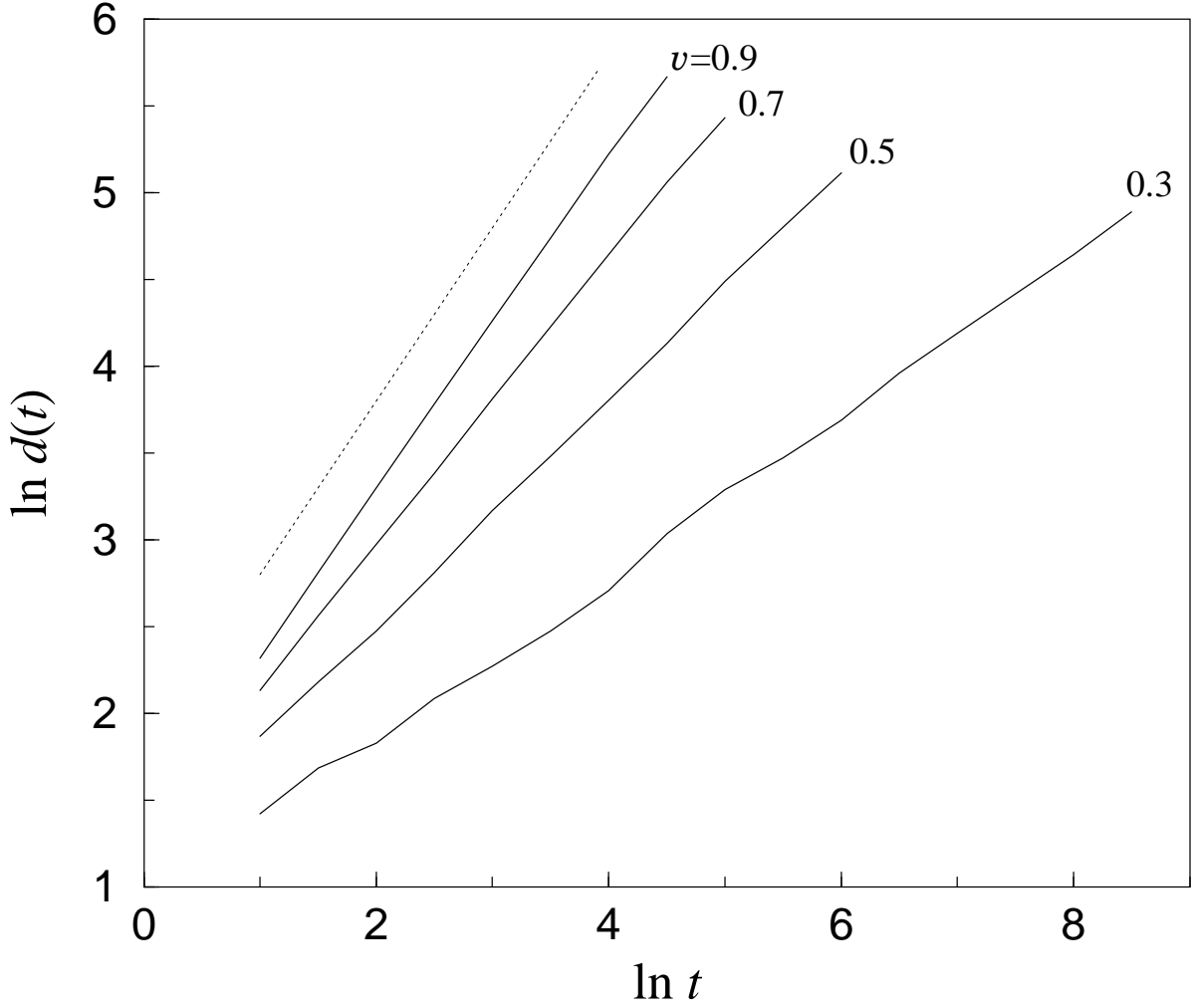,width=\columnwidth}}
 \caption{The mean square displacement $d(t)$ for the labyrinth tiling shows 
$d(t)\sim t^{-\beta}$ with $\beta=0.95$, $0.83$, $0.66$, and $0.47$ for 
$v=0.9$, $0.7$, $0.5$, and $0.3$, respectively. 
The dotted line corresponds to ballistic motion $d(t)\sim t$.\label{fig:Fig8}}
\end{figure}

\clearpage
\begin{figure} 
\centerline{\psfig{figure=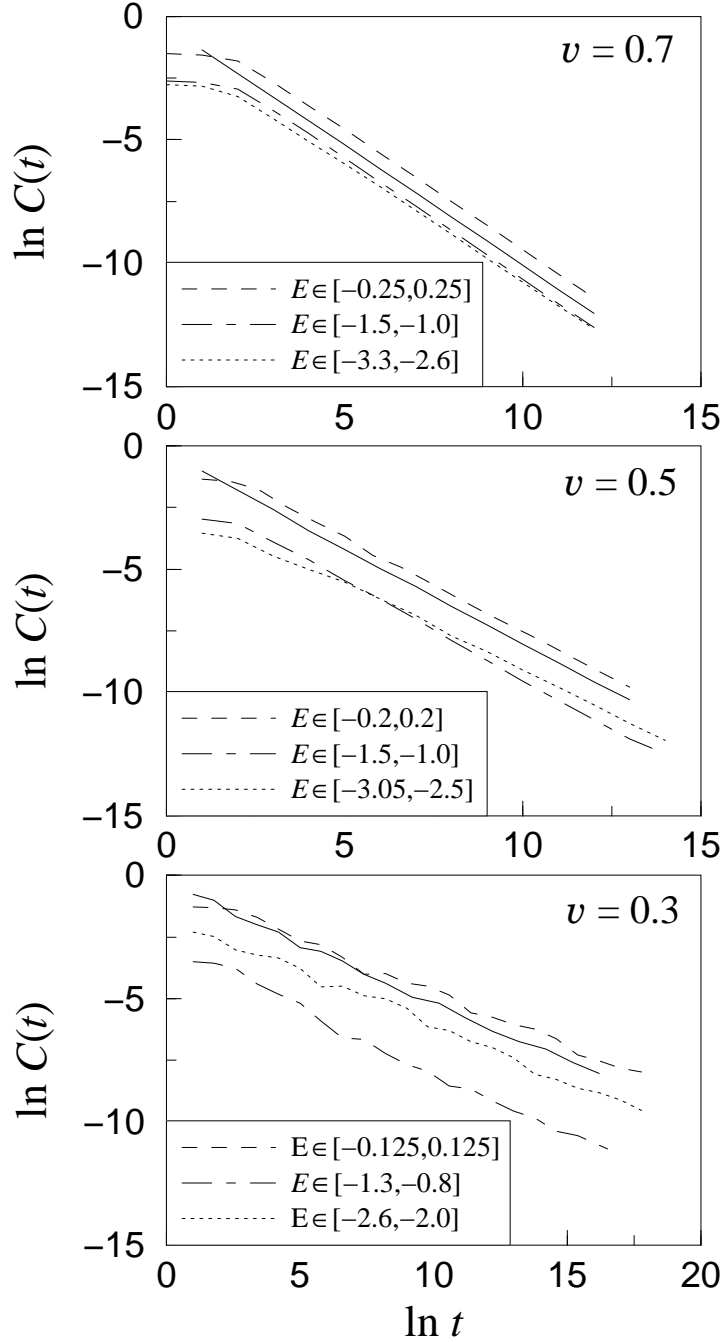,width=0.6\columnwidth}}
 \caption{The dependence of autocorrelation function $C(t)$ on various
 energy windows. The energy windows are arbitrarily chosen from the
 band center (dashed line), the band edge (dotted line) and between
 the band center and the band edge (dot-dashed line),
 respectively. The solid line is the result for the full spectrum.
\label{fig:Fig9}}
\end{figure}

\clearpage
\begin{figure} 
\centerline{\psfig{figure=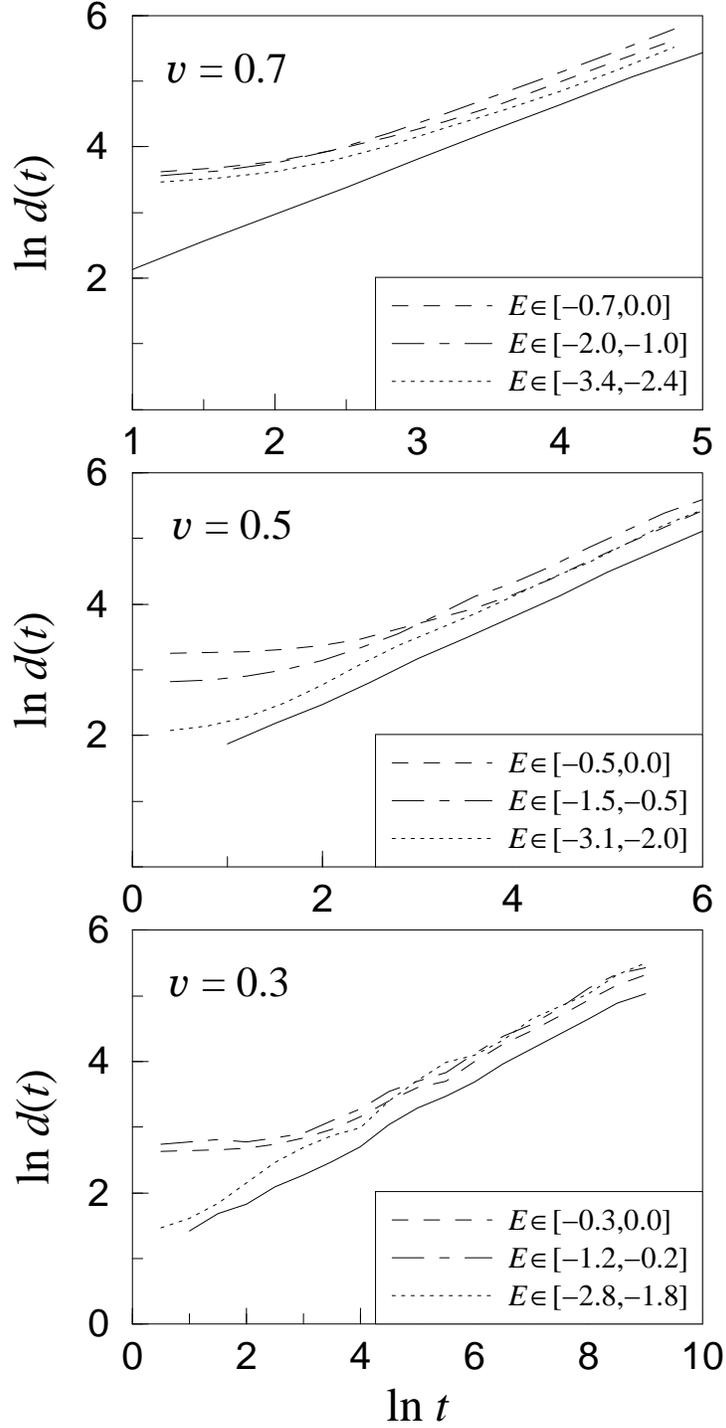,width=0.6\columnwidth}}
 \caption{The dependence of mean square displacement $d(t)$ on various
 energy windows. The energy windows are arbitrarily chosen from the
 band center (dashed line), the band edge (dotted line) and between
 the band center and the band edge (dot-dashed line),
 respectively. The solid line is the result for the full spectrum.
\label{fig:Fig10}}
\end{figure}

\clearpage
\begin{figure} 
\centerline{\psfig{figure=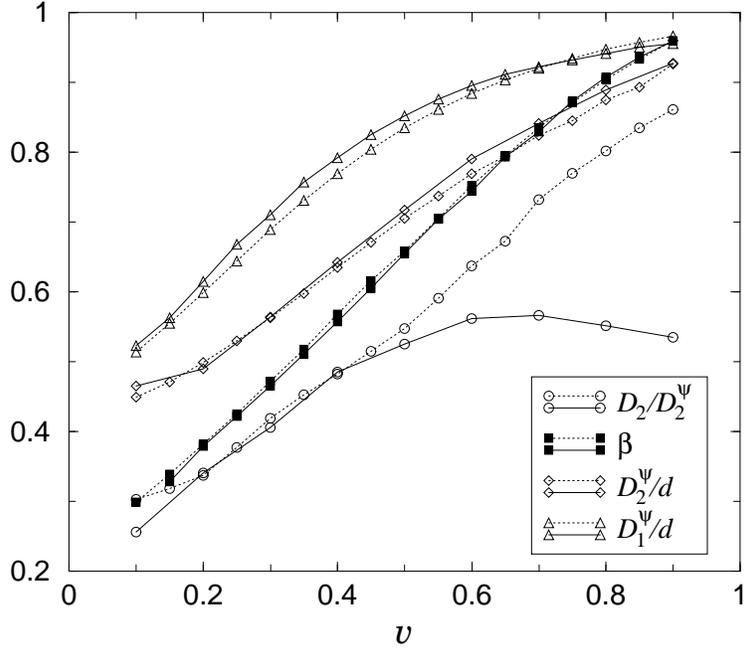,width=0.6\columnwidth}}
 \caption{The exponent $\beta$ compared to several quantities
 describing the multifractal properties of the energy spectra and
 wavefunctions. The lines are included to guide the eye; the dotted
 lines correspond to the results for the octonacci chain ($d=1$), the
 solid lines to the labyrinth tiling ($d=2$).\label{fig:Fig11}}
\end{figure}

\end{document}